# Entropy-Enthalpy Compensation May Be a Useful Interpretation Tool for Complex Systems Like Protein-DNA Complexes: An Appeal to Experimentalists

E. B . Starikov[1,2], B. Nordén[2]

[1] *Institute for Materials Science and Max Bergmann Center of Biomaterials, Dresden University of Technology, D-01062 Dresden, Germany,* E-mail: starikow@tfp.uni-karlsruhe.de

[2] *Department of Physical Chemistry, Chalmers University of Technology, SE-412 96 Gothenburg, Sweden*, E-mail: norden@chalmers.se

## Abstract

In various chemical systems enthalpy-entropy compensation (EEC) is a well-known rule of behavior, although the physical roots of it are still not completely understood. It has been frequently questioned whether EEC is a truly physical phenomenon or a coincidence due to trivial mathematical connections between statistical-mechanical parameters – or even simpler: A phantom effect resulting from the misinterpretation of experimental data. Here, we review EEC from a new standpoint using the notion of correlation which is essential for the method of factor analysis, but is not conventional in physics and chemistry. We conclude that the EEC may be rationalized in terms of hidden (not directly measurable with the help of the current experimental set-up) but physically real factors, implying a Carnot-cycle model in which a micro-phase transition (MPT) plays a crucial role. Examples of such MPTs underlying physically valid EEC should be typically cooperative processes in supramolecular aggregates, like changes of structured water at hydrophobic surfaces, conformational transitions upon ligand-biopolymer binding, and so on, so forth. The MPT notion could help rationalize the occurrence of EEC in connection with hydration and folding of proteins, enzymatic reactions, functioning of molecular motors, DNA de- and rehybridization, as well as similar phenomena.

In the recent papers [1,2], the effects of genetic variants of TGACGTCA DNA binding motif, as well as of successive C-terminal truncation of leucine zippers, on the energetics of DNA binding to bZIP domains in Jun transcription factor have been studied using analytical laser scattering in combination with isothermal titration calorimetry. The systematical study [1] reveals that the bZIP domains exhibit differential energetics in binding to DNA fragments containing single nucleotide variations within the TGACGTCA canonical motif. Further, it has been persuasively shown [2] that the successive C-terminal truncation of residues leading up to each signature leucine significantly compromises the binding of bZIP domains to the canonical DNA motif. Moreover, in these both works a valid enthalpy-entropy compensation (EEC) has been revealed.

When speaking of the EEC finding, the works [1,2] reference our paper [3], where we have suggested a new generalized model to rationalize the EEC phenomenon in terms of hidden, but physically real, factors implying a (real or imaginary) Carnot cycle in which some kind of micro-phase transition (MPT) plays a crucial role. But the authors [1,2] have not applied our model to their EEC data.

The present communication reports on the results of processing the EEC data of the works [1,2] with our model [3], as well as some further discussion on leucine zippers role in DNA binding.

Mathematically, the EEC can be expressed as a linear regression of enthalpy $H$ on entropy $S$, that is, $H = a*S + b$, where $a$ is the so-called "compensation temperature" and $b$ has energy dimension.

By fitting the relevant data from Table 1 in the work [1] with such an expression, we obtain a reliable and definite $H$-$S$ linear regression (see Fig. 1 and its legend for the details of the regression analysis), with the coefficients: $a$ = 305 K, $b$ = – 7.48 kcal/mol. Thus, on the $S$-$T$ diagram of the corresponding "imaginary/hidden Carnot cycle", the temperature is slightly going up from room temperature 298.15 K to 305 K, whereas the pertinent entropy difference is $\Delta S = b/a = -24.51$ cal/(mol K).

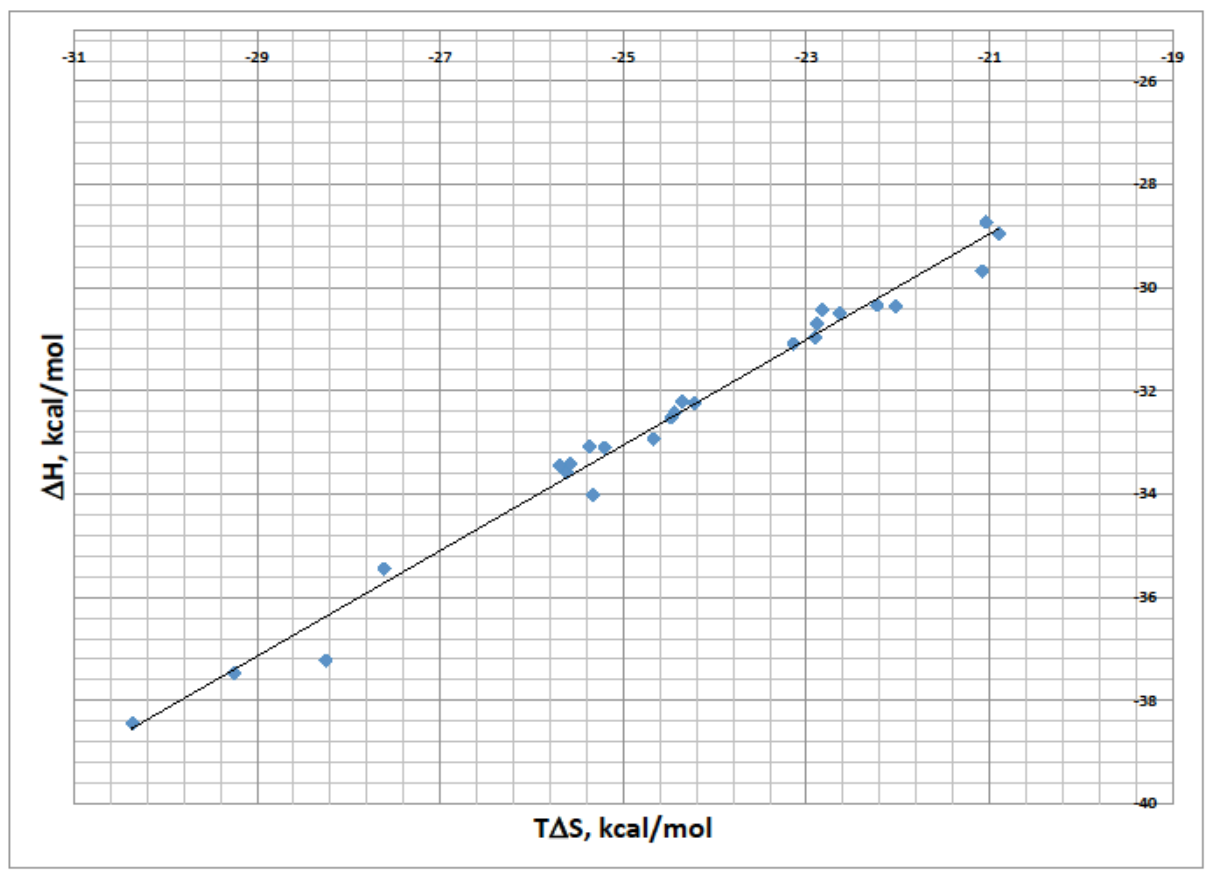

**Fig. 1**. Enthalpy-entropy plot, with 25 points of the experimental data [1], and its best linear fit. The regression equation for this plot is $\Delta H = 1.023*T*\Delta S - 7.475$, $T$ = 298.15 K, standard error for the slope is 0.027, standard error for the intercept is 0.658, the correlation coefficient is 0.992, its standard error is 0.026, residual sum of squares is 2.429. According to the conventional residual-based diagnostic tools, there is little evidence against the normality of the data set, whereas according to the F-statistic, there are little or no real evidences against the linearity of the plot. Therefore, we obtain a valid linear regression, with the slope greater or equal to 1.

Similarly, by fitting the relevant data from Table 2 in the work [2] with the same expression, we obtain a reliable and definite *H-S* linear regression (see Fig. 2 and its legend for the details of the regression analysis), with the coefficients: $a$ = 291 K, $b$ = – 8.91 kcal/mol. Hence, on the *S-T* diagram of the corresponding "imaginary/hidden Carnot cycle", the temperature is slightly going down from room temperature 298.15 K to 291 K, whereas the pertinent entropy difference is $\Delta S = b/a$ = – 30.62 cal/(mol K).

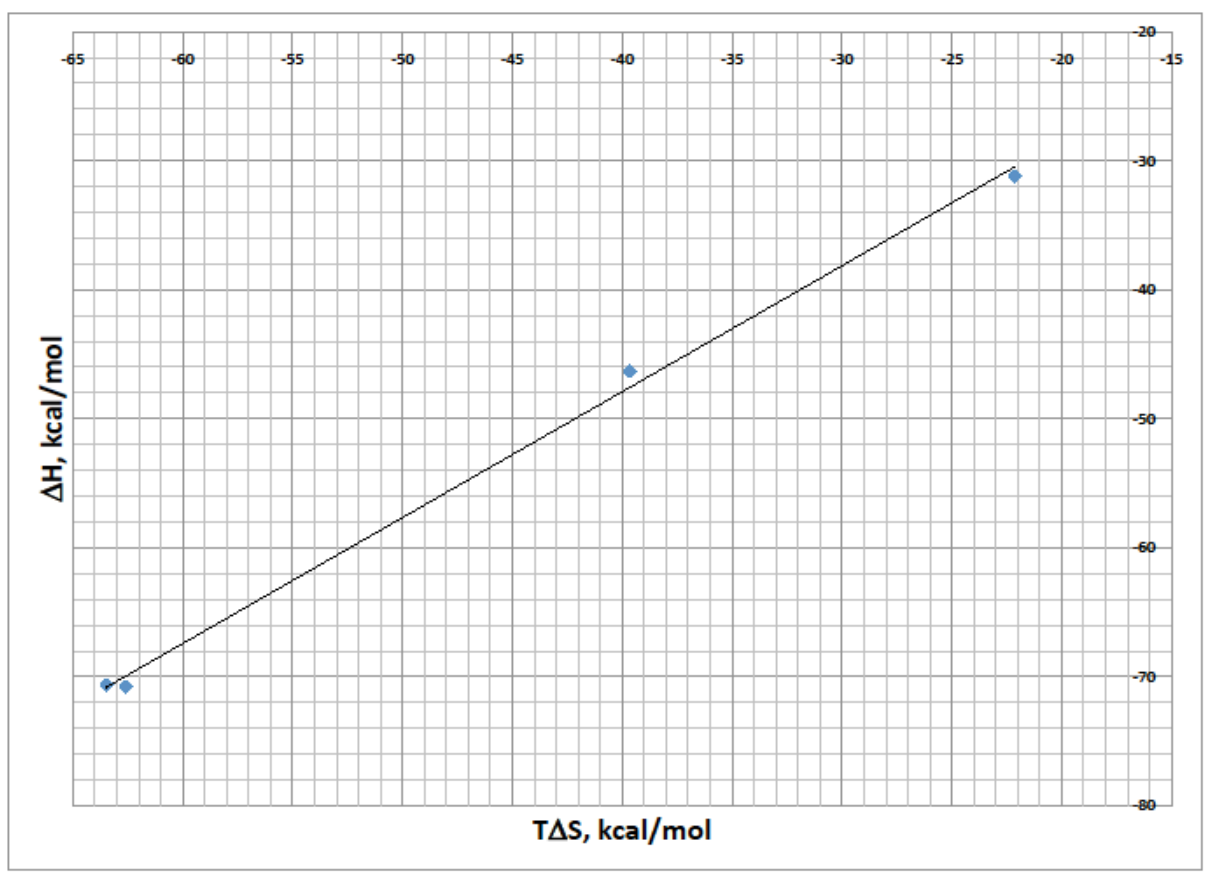

**Fig. 2**. Enthalpy-entropy plot, with 4 points of the experimental data [2], and its best linear fit. The regression equation for this plot is $\Delta H = 0.976*T*\Delta S – 8.907$, $T$ = 298.15 K, standard error for the slope is 0.034, standard error for the intercept is 1.681, the correlation coefficient is 0.999, its standard error is 0.034, residual sum of squares is 2.679. According to the conventional residual-based diagnostic tools, there is no evidence against the normality of the data set, whereas according to the F-statistic, there are little or no real evidences against the linearity of the plot. Therefore, we obtain a valid linear regression, with the slope less or equal to 1.

Hence, systematic modification of the nucleotide sequence in the DNA binding motif is corresponding to an "imaginary artificial heat pump" picture, whereas changing the length of the leucine zipper can thermodynamically be described as an "imaginary artificial refrigerator" model. By analogy with DNA-dye binding [3], the "refrigerator" effect means that the dynamics of the bZIP domain is *anti-correlated* with the DNA motions: The less motile the bonded bZIP ("cooling"), the more intensive are the motions of the DNA and its counterion-hydration shell ("heating"), and vice versa. Accordingly, the "heat pump" picture implies that the dynamics of the bZIP domain is *correlated* with the DNA motions: The more motile the bonded bZIP ("heating"), the more intensive are the motions of the DNA and its counterion-hydration shell ("heating"), and vice versa.

Furthermore, in analogy with DNA binding to diazapyrene cations carryng different electric charges [3], getting rid of the L5 heptad of the LZ, one omits two lysine residues, and therefore an electric charge of +2. Further, cutting out the L4 heptad, one omits one arginine and one glutamic acid residue, so that there is no electric charge change, whereas the L3 heptad contains no charge amino-acid residues at all. Finally, deleting the L2 or L1 heptads, one omits one lysine and one glutamic acid residue, or two lysine and two glutamic acid residues, respectively, and produces no electric charge change again. With this in mind, we can see that, along with diminishing their LZ chain length, the ΔL5, ΔL54 and ΔL543 mutants lose an electric charge of +2 in comparison to the wild type, and this situation is indeed more or less similar to what has taken place in the DNA-diazapyrene binding studies [4]. It is important to note here that DNA-diazapyrene binding can also be described as an "artificial refrigerator" [3].

Along with this, just modifying DNA nucleotide sequence does not influence the electrostatics of the DNA protein complex. Besides, the lengths of the corresponding DNA and protein fragments remain the same. This resembles the situation studied at binding of piperazinylcarbonyloxyalkyl ligands to DNA [5], which can be interpreted as an "artificial heat pump" as well [3]. Immobilized DNA together with the immobilized proper stretch of the protein that directly binds to it will be capable of contributing the full enthalpic factor to the over-all binding constant. At the same time, the effective transmission of the molecular dynamics from the immobilized DNA-protein complex to the neighboring leucine zipper will significantly decrease the corresponding entropic factor.

Hence, the DNA-bZIP binding ought to be a dynamical process with the complicated interplay between the electrostatic and hydrophobic factors, while DNA can be most effectively immobilized near the bZIP binding region when the LZ dynamics becomes more intensive. And, *vice versa*, to promote the bonded DNA motility/release, one would have to somehow constrain the LZ dynamics. Being immobilized through the dynamics of the leucine zipper, DNA fragment will in turn "tame" the dynamics of the very "binding active site" of the bZIP domain. Therefore, enthalpy-entropy compensation is, physically, not like a "potential barrier to be overcome", as the authors [1,2] conclude. Instead, it ought to be a generalized "smart method" used by biological macromolecules to achieve *both* enthalpic *and* entropic gains in one and the same process like molecular recognition, thus tremendously increasing the efficiency of the latter.

The above is only one example of how the approach published in [3] could successfully be applied to interpret systematical kinetic or equilibrium-thermodynamic studies of rather complicated processes. Whereas the work [3] gives thorough theoretical analysis of the EEC phenomenon, here we would like to present the simple interpretational algorithm:

1. Thorough and systematical experimental data on EEC should first be obtained (like in the works [1,2], for example). But mind that not every experimentally revealed EEC is a valid one ! Typically, one must use independent experimental approaches for one and the same specimen (set of specimens) to get enthalpy and entropy. If the latter both are obtained, say, as a result of the conventional Arrhenius or van't Hoff analyses, this is not a physically-chemically interpretable EEC.

2. The conventional linear regression of the experimental enthalpy on the experimental entropy data must be found in the standard way, to evaluate the *a* and *b* parameters. Then, the "Carnot entropic parameter", *b/a*, can be determined.

3. The results thus obtained can be interpreted using the own experimental data and the information known from the literature.

The algorithm in itself is pretty easy, with the third step being surely the most non-trivial one. But this should not constitute any "inviolable fortification" for the specialists in the respective fields.

To this end, it is very important to summarize the intrinsic difficulties, as well as the positive breakthroughs connected with employing the EEC concept in different fields of physical chemistry.

Specifically, first of all, we have to mention here the simple intrinsic relationship between the effect of solvent and temperature on the chromatographic retention in the reversed-phase-chromatography (RPC) that arises from the previously observed enthalpy-entropy compensation, which ought to be *an "extrathermodynamic" relationship*, as several author groups conclude, after the detailed statistical-mechanical analysis and careful, systematical experimental work [5].

Along with this, the EEC phenomenon has even been considered something which ought to be *overcome* (!) to achieve the proper molecular host-guest "binding affinity" [7]. On the other hand, the EEC can in principle be treated as a kind of driving force when studying protein folding and hydration [8,9] (especially, in dealing with the salt and osmolytic effects on the molecular-scale hydrophobic hydration and interactions [10]).

Still, as concerns molecular/macromolecular binding (or combined binding-folding/refolding) processes, the role of the EEC phenomenon is not just unambiguously "impeding", as one might immediately conclude after carefully reading, say, the work [7], as clearly demonstrated in the recent papers [1,2,11]. The EEC phenomenon is, furthermore, definitely relevant to enzymatic processes, to the interaction of amino-acid residues in proteins with the water of hydration, in particular, as well as – most probably – to the molecular/supramolecular-crowding-induced self-assembly, in general [12-14].

Interestingly, several recent works devoted to the EEC phenomenon completely support the standpoint that the EEC is of essential mechanistic significance for the processes involving the host-guest (supra)molecular binding, as well as the physical-chemical events triggered by the latter ones. And, along with all this, the EEC phenomenon ought to be deeply rooted in the thermodynamics [15-19].

Remarkably, all the most recent works are completely in line with the above conclusion, for they are demonstrating examples of the correct and successful usage of the EEC concept when trying to explain systematical experimental data obtainable in different fields of physical chemistry [20-23].

When investigating the physical-chemical significance of the enthalpy-entropy compensation principle, it is extremely important to find the detailed connection of the latter to the basics of thermodynamics, as well as to properly refine the sense of the entropy notion. And such studies are also underway in several groups all over the world, see, e. g., the recent works [24-28].

*Acknowledgement*: B. N. would like to acknowledge financial support from King Abdullah University of Science and Technology (KAUST).

**References**

[1] Seldeen, K. L.; McDonald, C. B.; Deegan, B. J.; Bhat, V.; Farooq A. DNA Plasticity Is a Key Determinant of the Energetics of Binding of Jun-Fos Heterodimeric Transcription Factor to Genetic Variants of TGACGTCA Motif. *Biochemistry* **2009** *48* 12213-12222.

[2] Seldeen, K. L.; McDonald, C. B.; Deegan, B. J.; Bhat, V.; Farooq A. Dissecting the role of leucine zippers in the binding of bZIP domains of Jun transcription factor to DNA. *Biochem. Biophys. Res. Comm.* **2010** *394* 1030-1035.

[3] Starikov, E. B.; Nordén, B. Enthalpy–entropy compensation: a phantom or something useful ? *J. Phys. Chem. B* **2007** *111* 14431–14435.

[4] Becker, H. C.; Nordén, B. DNA Binding Properties of 2,7-Diazapyrene and Its *N*-Methylated Cations Studied by Linear and Circular Dichroism Spectroscopy and Calorimetry. *J. Am. Chem. Soc.* **1997** *119* 5798-5803.

[5] (a) Becker, H. C.; Nordén, B. DNA Binding Mode and Sequence Specificity of Piperazinylcarbonyloxyethyl Derivatives of Anthracene and Pyrene. *J. Am. Chem. Soc.* **1999** *121* 11947-11952; (b) H. C. Becker, B. Nordén, DNA Binding Thermodynamics and Sequence Specificity of Chiral Piperazinecarbonyloxyalkyl Derivatives of Anthracene and Pyrene. *J. Am. Chem. Soc.* **2000** *122* 8344-8349.

[6] (a) Melander, W. R.; Chen, B.-K.; Horvath, C. Mobile Phase Effects in Reversed Phase Chromatography I. Concomitant Dependence of Retention on Column Temperature and Eluent Composition. *J. Chromatogr.* **1979** *185* 99-109; (b) Tan, L. Ch.; Carr, P. W. Extra-thermodynamic Relationships in Chromatography – Study of the Relationship Between the Slopes and Intercepts of Plots of ln k' vs. Mobile Phase Composition in Reversed-Phase Chromatography. *J. Chromatogr. A* **1993** *656* 521-535; (c) Poole S. K.; Kollie, Th. O.; Poole, C. F. Influence of Temperature on the Mechanism by Which Compounds are Retained in Gas-Liquid Chromatography. *J. Chromatogr. A* **1994** *664* 229-251; (d) Li, J.; Carr P. W. Extra-Thermodynamic Relationships in Chromatography. Enthalpy-Entropy Compensation in Gas Chromatography. *J. Chromatogr. A* **1994** *670* 105-116.

[7] Rekharsky, M. V.; Mori T.; Yang, Ch.; Young, H. K.; Selvapalam, N.; Kim, H.; Sobransingh, D.; Kaifer, A. E.; Liu, S.; Isaacs, L.; Chen, W.; Moghaddam, S.; Gilson, M. K.; Kim. K.; Inoue, Y. A synthetic host-guest system achieves avidin-biotin affinity by overcoming enthalpy-entropy compensation. *Proc. Natl. Acad. Sci. USA* **2007** *104* 20737-20742.

[8] Nickson, A. A.; Stoll, K. E; Clarke, J., Folding of a LysM Domain: Entropy-Enthalpy Compensation in the Transition State of an Ideal Two-state Folder. *J. Mol. Biol.* **2008** *380*

557–569.

[9] Leung, D. H.; Bergman, R. G.; Raymond, K. N., Enthalpy-Entropy Compensation Reveals Solvent Reorganization as a Driving Force for Supramolecular Encapsulation in Water. *J. Am. Chem. Soc*. **2008** *130* 2798-2805.

[10] Athawale, M. V.; Sarupria, S.; Garde Sh., Enthalpy-Entropy Contributions to Salt and Osmolyte Effects on Molecular-Scale Hydrophobic Hydration and Interactions. *J. Phys. Chem. B* **2008** *112* 5661-5670.

[11] Joynt, S.; Morillo, V.; Leng, F., Binding the Mammalian High Mobility Group Protein AT-hook 2 to AT-Rich Deoxyoligonucleotides: Enthalpy-Entropy Compensation. *Biophys. J.* **2009** *96* 4144–4152.

[12] Edwards, A. A.; Mason, J. M.; Clinch, K; Tyler, P. C.; Evans, G. B.; Schramm, V. L., Altered Enthalpy-Entropy Compensation in Picomolar Transition State Analogues of Human Purine Nucleoside Phosphorylase. *Biochemistry* **2009** *48* 5226–5238.

[13] Kurhe, D. N.; Dagade, D. H.; Jadhav, J. P.; Govindwar, S. P.; Patil, K. J., Studies of Enthalpy-Entropy Compensation, Partial Entropies, and Kirkwood-Buff Integrals for Aqueous Solutions of Glycine, L-Leucine, and Glycylglycine at 298.15º K. *J. Phys. Chem. B* **2009** 113, 16612–16621.

[14] Douglas, J. F.; Dudowicz, J.; Freed, K. F. Crowding Induced Self-Assembly and Enthalpy-Entropy Compensation. *Phys. Rev. Lett*. **2009** *103* 135701.

[15] Ward, J. M.; Gorenstein, N. M.; Tian, J.; Martin, S. F.; Post, C. B., Constraining Binding Hot Spots: NMR and Molecular Dynamics Simulations Provide a Structural Explanation for Enthalpy-Entropy Compensation in SH2-Ligand Binding. *J. Am. Chem. Soc*. **2010** *132* 11058-11070.

[16] Liang, F.; Cho, B. P., Enthalpy-Entropy Contribution to Carcinogen-Induced DNA Conformational Heterogeneity. *Biochemistry* **2010**, *49*, 259–266.

[17] Yu, H.; Rick, S. W., Free Energy, Entropy, and Enthalpy of a Water Molecule in Various Protein Environments. *J. Phys. Chem. B* **2010**, *114,* 11552–11560.

[18] Kocherbitov, V.; Arnebrant, Th., Hydration of Lysozyme: The Protein-Protein Interface and the Enthalpy-Entropy Compensation. *Langmuir* **2010** *26* 3918-3922.

[19] Mauro, J. C.; Loucks, R. J.; Sen, S., Heat capacity, enthalpy fluctuations, and configurational entropy in broken ergodic systems *J. Chem. Phys*. **2010** *133* 164503.

[20] Olsson, T. S.; Ladbury, J. E.; Pitt, W. R.; Williams M. A., Extent of enthalpy-entropy compensation in protein-ligand interactions. *Protein Sci*. **2011** *20* 1607-1618.

[21] Piguet, C., Enthalpy–entropy correlations as chemical guides to unravel self-assembly processes. *Dalton Trans*. **2011**, *40*, 8059-8071.

[22] Wilfong, E. M.; Kogiso, Y.; Muthukrishnan, S.; Kowatz, Th.; Du, Y.; Bowie, A.;

Naismith, J. H.; Hadad, Ch. M.; Toone, E. J.; Gustafson, T. L., A Multidisciplinary Approach to Probing Enthalpy-Entropy Compensation and the Interfacial Mobility Model. *J. Am. Chem. Soc.* **2011** *133* 11515–11523.

[23] Freed, K. F., Entropy-Enthalpy Compensation in Chemical Reactions and Adsorption: An Exactly Solvable Model. *J. Phys. Chem. B* **2011** *115* 1689–1692.

[24] Freire, E., Do Enthalpy and Entropy Distinguish First in Class from Best in Class ? *Drug Discov. Today* **2008** *13* 869–874.

[25] Lambert, F. L.; Leff, H. S.; The Correlation of Standard Entropy with Enthalpy Supplied from 0º to 298.15º K. *J. Chem. Educ.* **2009** *86* 94–98.

[26] Starikov, E. B., Many Faces of Entropy or Bayesian Statistical Mechanics. *ChemPhysChem* **2010** *11* 3387–3394.

[27] Chirikjian, G. S., Modeling Loop Entropy. *Meth. Enzymol.* **2011** *487* 99–132.

[28] Vicente, R.; Wibral, M.; Lindner, M.; Pipa. G., Transfer Entropy – a Model-Free Measure of Effective Connectivity for the neurosciences. *J. Comput. Neurosci.* **2011** *30* 45–67.